\documentclass[10pt,twocolumn,twoside]{IEEEtran}
\usepackage{amsmath,array,amssymb,graphicx,color}
\usepackage{cite}
\usepackage{url}
\usepackage{amsthm}
\usepackage{colortbl} % rowcolor

\def\thline{\noalign{\hrule height 1pt}}

\newcommand{\mb}[1]{\mathbf{#1}} % math bold
\newcommand{\ul}[1]{\underline{#1}} % under line
\begin{document}

\title{
Lightning-Fast Dual-Layer Lossless Coding for\\Radiance Format High Dynamic Range Images
}
\author{
Taizo~Suzuki,~\IEEEmembership{Senior Member,~IEEE},
Sara~Yukikata,
Kai~Yang,
and Taichi~Yoshida,~\IEEEmembership{Member,~IEEE}
\thanks{This work was supported by JSPS Grant-in-Aid for Scientific Research (C) under Grant 22K04084.}
\thanks{T. Suzuki is with the Institute of Engineering, Information and Systems, University of Tsukuba, Ibaraki 305-8573, Japan (e-mail: taizo@cs.tsukuba.ac.jp).}
\thanks{S. Yukikata is with Sky Co., Ltd., Tokyo 108-0075, Japan (e-mail: yukikata@wmp.cs.tsukuba.ac.jp).}
\thanks{K. Yang is with Da-Jiang Innovations Science and Technology Japan Co., Ltd., Tokyo 108-0075, Japan (e-mail: roarkyang@gmail.com).}
\thanks{T. Yoshida is with the Department of Computer and Network Engineering, Graduate School of Informatics and Engineering, The University of Electro-Communications, Tokyo 182-8585, Japan (e-mail: t-yoshida@uec.ac.jp).}
}

\maketitle

%%%%%%%%%%%%%%%%%%%%%%%%%%%%%%%%%%%%%%%%%%%%%%%%%%
\begin{abstract}
%%%%%%%%%%%%%%%%%%%%%%%%%%%%%%%%%%%%%%%%%%%%%%%%%%
This paper proposes a fast dual-layer lossless coding for high dynamic range images (HDRIs) in the Radiance format.
The coding, which consists of a base layer and a lossless enhancement layer, provides a standard dynamic range image (SDRI) without requiring an additional algorithm at the decoder and can losslessly decode the HDRI by adding the residual signals (residuals) between the HDRI and SDRI to the SDRI, if desired.
To suppress the dynamic range of the residuals in the enhancement layer, the coding directly uses the mantissa and exponent information from the Radiance format.
To further reduce the residual energy, each mantissa is modeled (estimated) as a linear function, i.e., a simple linear regression, of the encoded-decoded SDRI in each region with the same exponent.
This is called simple linear regressive mantissa estimator.
Experimental results show that, compared with existing methods, our coding reduces the average bitrate by approximately $1.57$--$6.68$ \% and significantly reduces the average encoder implementation time by approximately $87.13$--$98.96$ \%.
\end{abstract}
%%%%%%%%%%%%%%%%%%%%%%%%%%%%%%%%%%%%%%%%%%%%%%%%%%
%\IEEEpeerreviewmaketitle
%%
%%%%%%%%%%%%%%%%%%%%%%%%%%%%%%%%%%%%%%%%%%%%%%%%%%
\section{Introduction}
%%%%%%%%%%%%%%%%%%%%%%%%%%%%%%%%%%%%%%%%%%%%%%%%%%
%% HDRI coding
\IEEEPARstart{R}{ADIANCE} and OpenEXR are common formats for high dynamic range image (HDRI) coding and are expressed in terms of floating points (floats) with mantissa and exponent information~\cite{Ward1991GGII,Kainz2003SIGGRAPH}.
To make a high-precision standard dynamic range image (SDRI) from these formats, traditional displays, i.e., decoders, require a tone mapping operator (TMO).
Considering the development costs and the market penetration, new codecs should not use the TMO at the decoder and should be able to use existing coders, such as JPEG~\cite{Wallace1992TCE}, directly.
Logarithmic methods~\cite{Xu2005CGA,Motra2010ICIP,Garbas2011ICASSP} logarithmically convert floats to integers and then encode HDRIs in accordance with standards such as JPEG 2000~\cite{Skodras2001SPM} and AVC~\cite{Wiegand2003IEEE}, but do not guarantee the reversibility between the encoding and decoding.
Similarly, standards such as HEVC~\cite{Francois2016TCSVT} and VVC~\cite{Bross2021TCSVT} and their improvements~\cite{Lee2020Access,Zhou2020TIP,Zhou2020TCSVT,Liu2022TCSVT,Zhou2022TIP} support HDRIs, but again do not guarantee the reversibility.

%% Dual-layer coding
Dual-layer coding for HDRIs~\cite{Boschetti2010ICIP,Banterle2012SPIE,Mai2011TIP,Watanabe2018IEICE,Wei2018TCSVTL} is another popular form of HDRI coding; for instance, it is used in the JPEG XT HDRI coding standard~\cite{Richter2013PCS}.
The encoder consists of a base layer for encoding a tone-mapped SDRI and an enhancement layer for encoding the residual signals (residuals) between the HDRI and SDRI.
It has two good properties, {\it image selectivity} and {\it backward compatibility}.
Boschetti et al. achieved a scalable coding for HDRIs by splitting an HDRI into a pseudo-exponent image like a grayscale image and a pseudo-mantissa like an SDRI~\cite{Boschetti2010ICIP}.
Banterle et al. employed a segmentation map to avoid seams or halos at the boundary of each segmented zone due to the TMO~\cite{Banterle2012SPIE}.
Mai et al. focused on optimizing the TMO at the cost of visual quality of the SDRIs~\cite{Mai2011TIP}.
Watanabe et al. used the histogram packing technique~\cite{Pinho2002SPL} to enhance the sparseness of the histogram values in the enhancement layer of the JPEG XT framework~\cite{Watanabe2018IEICE}.
Wei et al. proposed a two-layer local inverse TMO (ITMO) using a fast global edge-preserving smoothing technique~\cite{Wei2018TCSVTL}.
However, since Boschetti et al.'s and Banterle et al.'s methods~\cite{Boschetti2010ICIP,Banterle2012SPIE} were originally irreversible and the others~\cite{Mai2011TIP,Watanabe2018IEICE,Wei2018TCSVTL} use the logarithmic method~\cite{Xu2005CGA}, they are not theoretically reversible.

%% Dual-layer "lossless" coding
Although irreversible (lossy) coding is commonly believed a desirable trade-off for HDR transmission, we believe that dual-layer ``reversible (lossless)'' coding has the potential to become the de facto standard in the future, as it has a third good property, i.e., {\it reversibility}~\cite{Iwahashi2015JASP,Yoshida2018IEICE}.
Thought this property, it is possible to obtain various SDRIs even after encoding and to utilize many applications of HDRIs.
Whilst any dual-layer coding can offer image selectivity and backward compatibility, the dual-layer lossless coding must have forward/inverse reversible integer converters and coders in the enhancement layer.
Here, Iwahashi et al.'s method (basic method)~\cite{Iwahashi2015JASP} with a reversible integer converter is unable to produce highly compressed HDRIs due to high residual energy.
Yoshida et al. reduced the residual energy by employing range compression and an adaptive ITMO while using the integer converter of JPEG XR to guarantee reversibility~\cite{Yoshida2018IEICE}.
The drawback of Yoshida et al.'s method is its high computational cost in encoding and decoding.
For practical applications, the computational cost in the coder should be as low as possible.

%% proposed method
This study presents a ``lightning-fast'' dual-layer lossless coding for HDRIs in the Radiance format (Radiance HDRIs).
Unlike existing methods that handle high dynamic range signals in the enhancement layer, our approach {\it directly} uses mantissa and exponent information, which has low dynamic range.\footnote{Note that Boschetti et al. used newly estimated pseudo-mantissa/exponent information (images) in \cite{Boschetti2010ICIP} and could not guarantee reversibility.}
However, it is clear that the residual energy between the original mantissa, not the HDRI as it is, and the encoded-decoded SDRI, which is not to be modified any more, is rather high, which leads to poor coding performance.
To deal with this problem, we generate an estimated mantissa, which is modeled as a linear function, i.e., a simple linear regression, of the SDRI.
Note that the original mantissa has obvious discontinuities between regions with different exponents and the non-modified SDRI causes more discontinuities in the residuals.
In order to prevent a coding performance degradation due to the discontinuities, we estimate the mantissa in each region with the same exponent.
The proposed simple linear regressive mantissa estimator (SLRME) does not require any ITMO or complex estimation such as iterative optimization and learning, so it is very fast.
The parameters calculated in the encoder are transmitted to the decoder and are not recalculated afterwards.
The experimental results demonstrate that it outperformed existing methods in terms of bitrate, while significantly reducing the encoder implementation time.

%% Yang et al.'s method
A preliminary version of this study was presented in \cite{Yang2020APSIPA}, where we described an incomplete global approach for the mantissa estimator.
In this paper, we developed a lightning-fast local approach that achieves higher coding performance and significantly reduces the encoder implementation time.

{\it Notation:}
In what follows, $\cdot(p)$ denotes the $p$-th value and $\cdot_E$ denotes the values according to the exponent $E$.

    \begin{figure}[t]
        \centering
        \includegraphics[scale=0.4,keepaspectratio=true]{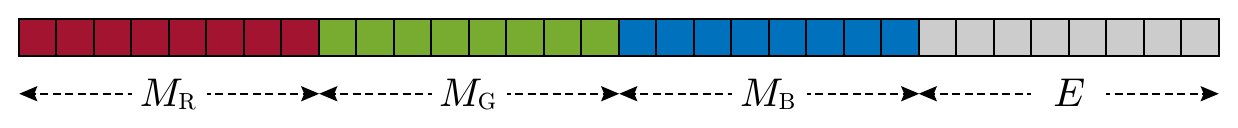}
        \caption{Radiance format (a square means a bit).}
        \label{Fig_Rad}
    \end{figure}

    \begin{figure*}[t]
        \centering
        \begin{tabular}{c}
        \includegraphics[scale=0.4,keepaspectratio=true]{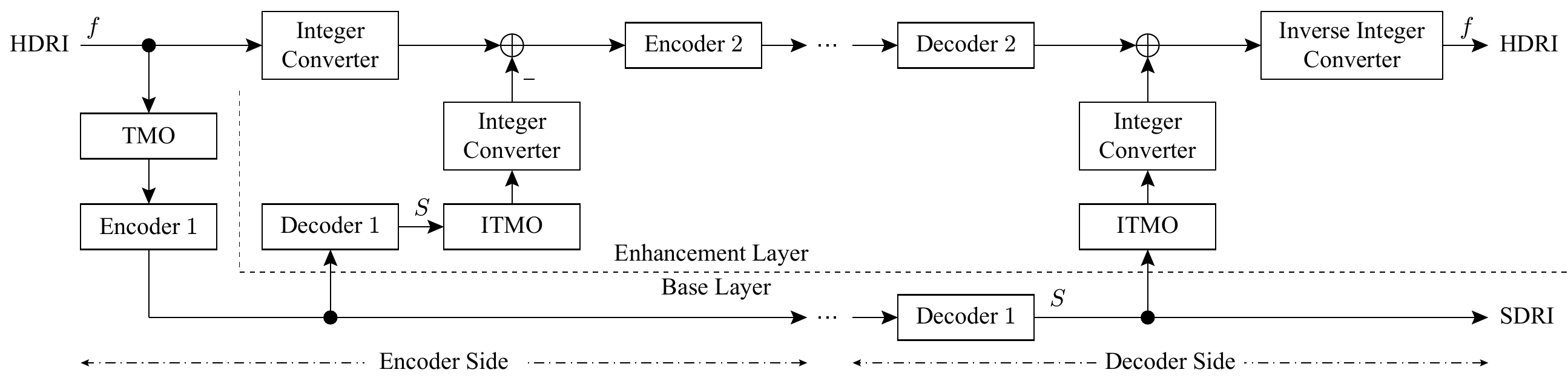} \\
        ~\\
        \includegraphics[scale=0.4,keepaspectratio=true]{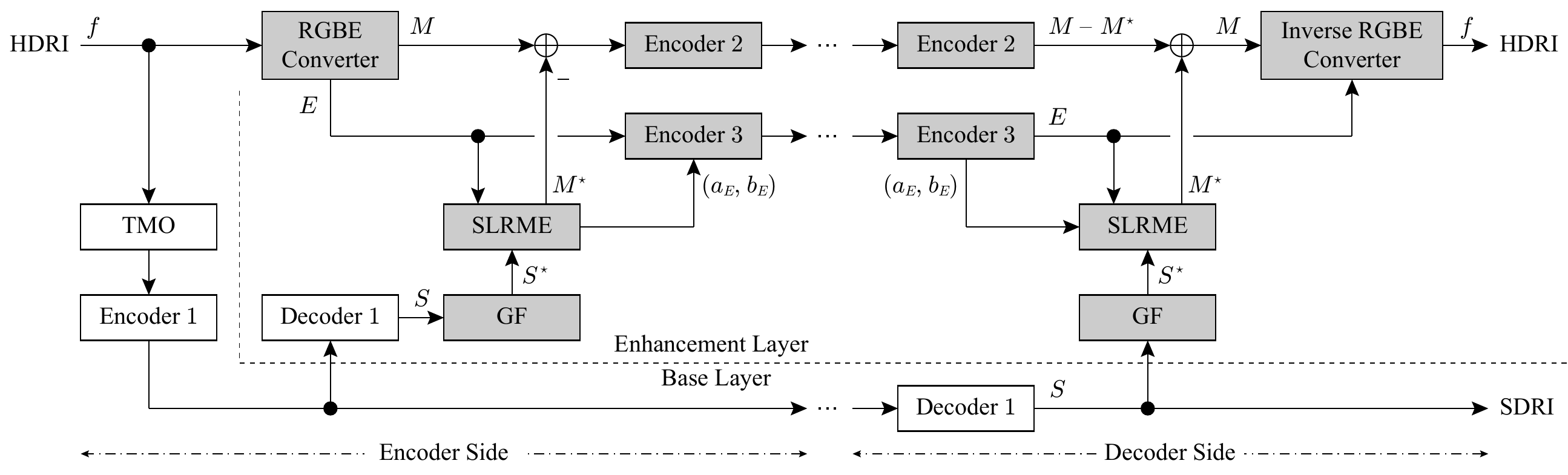}
        \end{tabular}
        \caption{Flows of dual-layer lossless coding: (top) basic procedure and (bottom) our procedure.}
        \label{Fig_Flow}
    \end{figure*}

%%%%%%%%%%%%%%%%%%%%%%%%%%%%%%%%%%%%%%%%%%%%%%%%%%
\section{Review and Definitions}
%%%%%%%%%%%%%%%%%%%%%%%%%%%%%%%%%%%%%%%%%%%%%%%%%%
%%%%%%%%%%%%%%%%%%%%%%%%%%%%%%
\subsection{Radiance Format}
%%%%%%%%%%%%%%%%%%%%%%%%%%%%%%
The Radiance format (`.hdr')~\cite{Ward1991GGII} is a widely used file format for representing HDRIs.
Each pixel is represented by assigning an $8$-bit integer to each of three mantissas $M\in\{M_\text{R},M_\text{G},M_\text{B}\}$, where $M_\times$ is the mantissa for each color, and an exponent $E$, resulting in a total of $8\times 4=32$ bits/pixel (bpp) as illustrated in Fig.~\ref{Fig_Rad}.
The format eliminates the sign bit and combines the exponent part into one value that does not depend on the mantissa, unlike the OpenEXR format~\cite{Kainz2003SIGGRAPH}.
The RGBE converter, which maps $16$-bit floats $f\in\{f_\text{R},f_\text{G},f_\text{B}\}$, where $f_\times$ is the float for each color, to $8$-bit integers $M$ and $E$, is defined as
    \begin{align}
        E(p)
        &
        = \left\lceil \log_2(\max(f(p))) + 128\right\rceil,
        \label{Eqs_E}
        \\
        M(p)
        &
        = \left\lfloor\frac{256 \cdot f(p)}{2^{E(p)-128}}\right\rfloor
        %= \left\lfloor 2^{136-E(p)} \cdot f(p) \right\rfloor
        .
        \label{Eqs_M}
    \end{align}
The inverse conversion, which maps $M$ and $E$ to $f$, is defined as
    \begin{align}
        f(p)
        = \frac{M(p) + 0.5}{256} \cdot 2^{E(p)-128}
        .
        \label{Eqs_f}
    \end{align}
The forward and inverse RGBE converters are perfectly reversible.

%%%%%%%%%%%%%%%%%%%%%%%%%%%%%%
\subsection{Dual-Layer Lossless Coding}
%%%%%%%%%%%%%%%%%%%%%%%%%%%%%%
The format has two layers, a base layer and an enhancement layer, as illustrated at the top of Fig.~\ref{Fig_Flow}.
This dual-layer coding has two good properties~\cite{Boschetti2010ICIP,Banterle2012SPIE,Mai2011TIP,Watanabe2018IEICE,Wei2018TCSVTL}:
    \begin{enumerate}
        \item Image selectivity: it provides both HDRIs and SDRIs,
        \item Backward compatibility: existing coders can be employed without having to modify them.
    \end{enumerate}
In the base layer, the original HDRI is first converted into an SDRI by applying an arbitrary TMO at the encoder and the SDRI can be directly reconstructed without any TMO at the decoder.
In the enhancement layer, the residuals between the original HDRI and another HDRI, generated by applying an arbitrary ITMO to the encoded-decoded SDRI, are encoded at the encoder and the HDRI is reconstructed at the decoder by applying the inverse operations.
When the forward/inverse integer converters and coders in the enhancement layer are reversible, the dual-layer coding has the third good property:
    \begin{enumerate}
        \setcounter{enumi}{2}
        \item Reversibility: It can reconstruct the HDRIs losslessly.
    \end{enumerate}
This is called dual-layer lossless coding~\cite{Iwahashi2015JASP,Yoshida2018IEICE,Yang2020APSIPA}.

%%%%%%%%%%%%%%%%%%%%%%%%%%%%%%
\subsection{Simple Linear Regression for Image Processing}
%%%%%%%%%%%%%%%%%%%%%%%%%%%%%%
Simple linear regression is a standard low-computation method for data analyses.
It linearly estimates an objective variable $Y$ from an explanatory variable $X$ using two parameters $a$ and $b$, as follows:
    \begin{align}
        Y(p) = a \cdot X(p) + b
        .
        \label{Eqs_reg}
    \end{align}
Since decreasing the estimation error means increasing the estimation accuracy, one calculates
    \begin{align}
        F(a,b)
        =
        \sum_{p=1}^{\mathcal{N}}
        |
        \underbrace{Y(p) - (a \cdot X(p) + b)}_{\text{estimation error}}
        |^2
        ,
        \label{Eqs_predErr}
    \end{align}
where $\mathcal{N}$ is the total number of $p$.
When each of the two equations obtained by partially differentiating $F(a,b)$ in \eqref{Eqs_predErr} with respect to $a$ and $b$ is set to $0$, $a$ and $b$ are uniquely determined minimum solutions, as follows:
    \begin{align}
        a
        &=
        \frac
        {\mathcal{N}\cdot\sum_p{(X(p) \cdot Y(p))} - \sum_p{X(p)}\cdot\sum_p{Y(p)}}
        {\mathcal{N}\cdot\sum_p{(X(p))^2} - (\sum_p{X(p)})^2}
        ,
        \label{Eqs_Alpha}
        \\
        b
        &=
        \frac
        {\sum_p{Y(p)} - a \cdot \sum_p{X(p)}}
        {\mathcal{N}}
        .
        \label{Eqs_Beta}
    \end{align}
Because of its simplicity, this sort of regression is often applied to images under certain conditions, such as guided filter~\cite{He2013TPAMI} and chroma from luma (CfL) prediction in AV1~\cite{Trudeau2018DCC}.

%%%%%%%%%%%%%%%%%%%%%%%%%%%%%%%%%%%%%%%%%%%%%%%%%%
\section{Our Coding Method}
%%%%%%%%%%%%%%%%%%%%%%%%%%%%%%%%%%%%%%%%%%%%%%%%%%
The lightning-fast dual-layer lossless coding for Radiance HDRIs is illustrated at the bottom of Fig.~\ref{Fig_Flow}.
We can see that the difference between the basic procedure at the top of Fig.~\ref{Fig_Flow} and the one at the bottom of Fig.~\ref{Fig_Flow} is only in the enhancement layer.

%%%%%%%%%%%%%%%%%%%%%%%%%%%%%%
\subsection{Incorporating RGBE Converter}
%%%%%%%%%%%%%%%%%%%%%%%%%%%%%%
Our coding incorporates the RGBE converter and the inverse in \eqref{Eqs_E}-\eqref{Eqs_f} into the forward/inverse integer converters at the upper line of the basic procedure.
Direct use of the mantissa and exponent reduces the dynamic range of the information handled in the enhancement layer and may increase compression efficiency.
However, since the residual energy between the original mantissa $M$ and the encoded-decoded SDRI $S$ is rather high if $S$ is not to be modified any more, the compression efficiency will not be as high as expected.
Here, existing codecs work well when the residuals are smooth and close to the color gray.

%%%%%%%%%%%%%%%%%%%%%%%%%%%%%%
\subsection{Simple Linear Regressive Mantissa Estimator}
\begin{figure*}[t]
        \centering
        \includegraphics[scale=0.35,keepaspectratio=true]{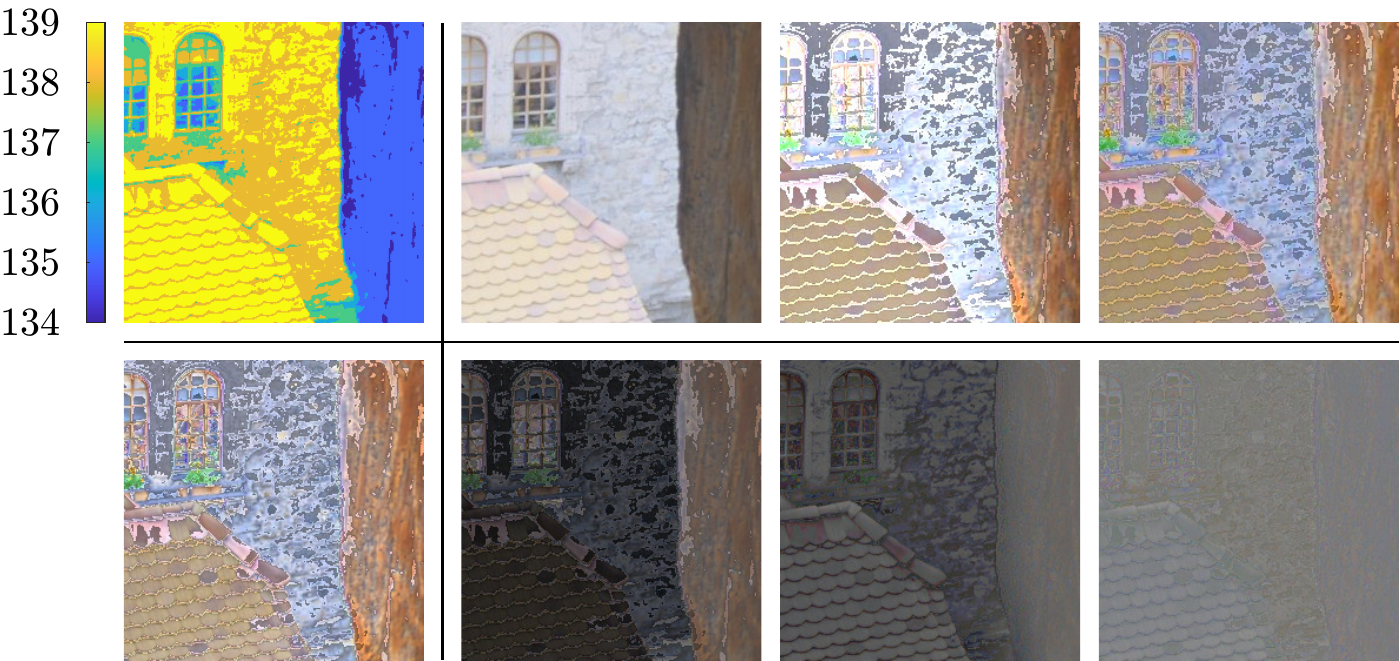}~~~% \\
        %~\\
        \includegraphics[scale=0.35,keepaspectratio=true]{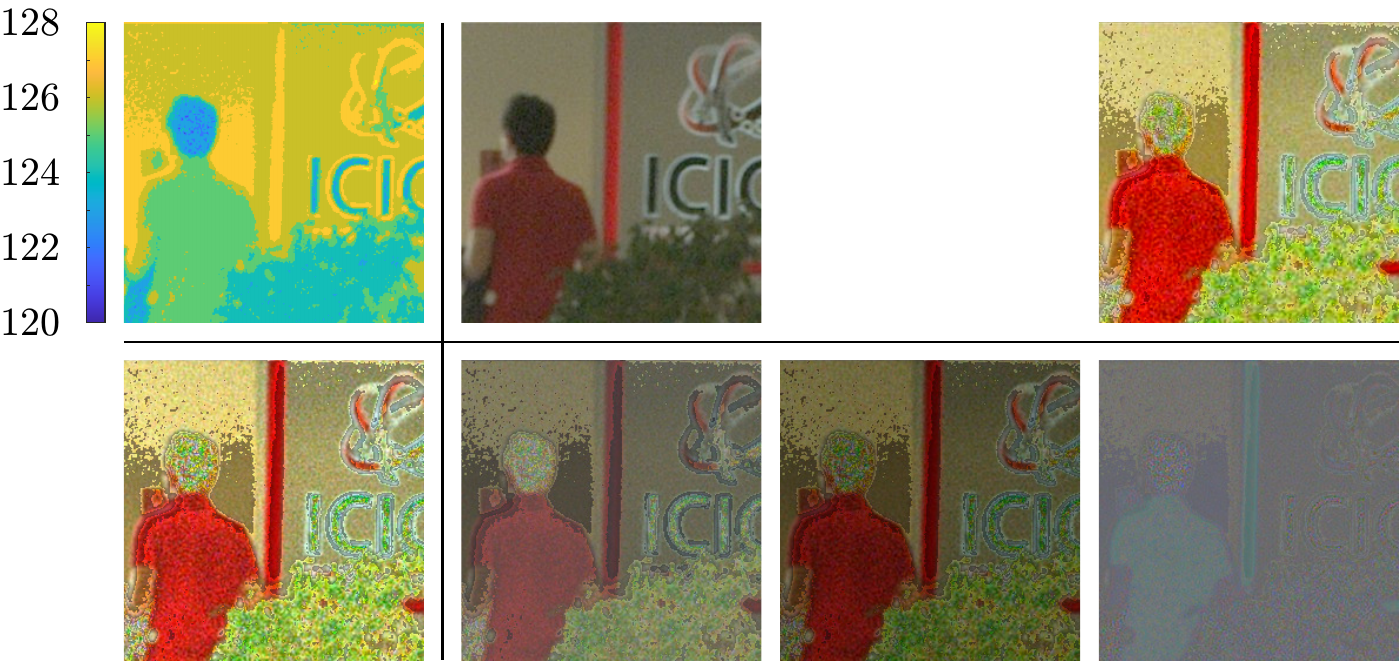}
        \caption{Differences between mantissas (when $Q=85$):
        (left) part of {\it C23} in \cite{Nemoto2015VPQM},
        (right) part of {\it ICICS} in \cite{Dehkordi2014QSHINE},
        (1st rows)
        $E$, $S$, $M^\prime$, and $M^\star$,
        and
        (2nd rows)
        $M$, $M-S$, $M-M^\prime$, and $M-M^\star$, where $E$ has been pseudo-colored and $M^\prime$ is the mantissa estimated by Yang et al.'s method~\cite{Yang2020APSIPA}.}
        \label{Fig_DiffMantissa}
    \end{figure*}

%%%%%%%%%%%%%%%%%%%%%%%%%%%%%%
We developed SLRME with the goal of making the residuals smooth and gray ones for good coding.
Here, SLRME estimates the mantissa from the encoded-decoded SDRI at a low computational cost.
Since the mantissa has discontinuities between regions with different $E$s and simple linear regression works well in the narrowest-possible area, SLRME uses simple linear regression to estimate $M$ in each region with the same $E$.
Also, since irreversible degradation of $S$ in the base layer may adversely affect coding performance, the method slightly restores $S$ at the expense of a small computation; specifically, it applies a Gaussian filter (GF) as a preprocessing to $S$.\footnote{Any restoration technique can be employed if its cost is low.}
Consequently, it makes use of the estimated mantissa $M^\star$, which is modeled as a simple linear regression of $S$, as follows:
    \begin{align}
        M^\star(q)
        =
        \text{round}(
        a_E \cdot S^\star(q) + b_E
        )_{[0, 255]}
        ,
        \label{Eqs_MstarL}
    \end{align}
where $q$ means the index determined for each $E$, $S^\star$ is the $S$ restored with the GF, and $\text{round}(\cdot)_{[0, 255]}$ denotes a rounding operation to $[0, 255]\in\mathbb{N}$.
In accordance with \eqref{Eqs_Alpha} and \eqref{Eqs_Beta}, $a_E$ and $b_E$ are uniquely determined, i.e., determined without requiring iterative optimization and learning, in each region with the same $E$, as follows:
    \begin{align}
        a_E
        &=
        \frac{
            \mathcal{N}_E\cdot
            \sum_{q}{(S^\star(q) \cdot M(q))}
            -
            \sum_{q}{S^\star(q)} \cdot \sum_{q}{M(q)}
        }{
            \mathcal{N}_E\cdot
            \sum_{q}{(S^\star(q))^2}
            -
            (\sum_{q}{S^\star(q)})^2
        }
        ,
        \\
        b_E
        &=
        \frac{
            \sum_{q}{M(q)}
            -
            a_E \cdot
            \sum_{q}{S^\star(q)}
        }{
            \mathcal{N}_E
        }
        ,
    \end{align}
where $\mathcal{N}_E$ is the total number of $q$.
As a result, it generates residuals, which are clearly smoother and closer to gray than those produced by Yang et al.'s method~\cite{Yang2020APSIPA}, as shown in Fig.~\ref{Fig_DiffMantissa}.
In particular, Yang et al.'s method could not estimate the mantissa in {\it ICICS} at all.

%%%%%%%%%%%%%%%%%%%%%%%%%%%%%%
\subsection{Influence on Coding Efficiency by Parameters}
%%%%%%%%%%%%%%%%%%%%%%%%%%%%%%
To provide both HDRIs and SDRIs, our dual-layer lossless coding transmits the SDRI, the residuals between $M$ and $M^\star$, and the parameters $(a_E,b_E)$ to the decoder.
Each of $(a_E,b_E)$ has only a $32$-bit value and in the experiments, there were at most 23 such parameters, making them of negligible impact on coding performance.
It is also computationally inexpensive, as the decoder does not have to recompute the parameters.

%%%%%%%%%%%%%%%%%%%%%%%%%%%%%%%%%%%%%%%%%%%%%%%%%%
\section{Experiments on Dual-Layer Lossless Coding}
%%%%%%%%%%%%%%%%%%%%%%%%%%%%%%%%%%%%%%%%%%%%%%%%%%
    \begin{table}[t]
        \centering
        \caption{Experimental bitrates [bpp], average bitrates [bpp], and average encoder implementation times [sec] (when $Q=85$).}
        \begin{tabular}{c|cc|cc}
        \thline
        Images & Yoshida+ & Yang+ & \multicolumn{2}{c}{Prop.} \\
        \cite{Nemoto2015VPQM,Dehkordi2014QSHINE} & \cite{Yoshida2018IEICE} & \cite{Yang2020APSIPA} & w/o SLRME & with SLRME \\
        \hline
        {\it C05} & $18.56$      & $\ul{17.41}$ & $18.47$      & $\mb{17.36}$ \\
        %{\it C08} & $16.68$      & $\ul{16.35}$ & $16.88$      & $\mb{15.95}$ \\
        %{\it C11} & $18.75$      & $\ul{17.87}$ & $18.81$      & $\mb{17.73}$ \\
        {\it C14} & $11.49$      & $\ul{11.02}$ & $\ul{11.02}$ & $\mb{10.72}$ \\
        %{\it C17} & $13.74$      & $\ul{11.89}$ & $12.31$      & $\mb{11.84}$ \\
        %{\it C20} & $12.55$      & $12.10$      & $\ul{11.73}$ & $\mb{11.61}$ \\
        {\it C23} & $15.86$      & $\ul{15.75}$ & $16.51$      & $\mb{15.14}$ \\
        %{\it C26} & $13.74$      & $\ul{13.39}$ & $13.49$      & $\mb{12.95}$ \\
        %{\it C29} & $13.90$      & $\ul{13.17}$ & $14.28$      & $\mb{13.14}$ \\
        {\it C32} & $13.30$      & $\ul{12.50}$ & $12.73$      & $\mb{12.22}$ \\
        %{\it C35} & $17.76$      & $\ul{15.72}$ & $16.32$      & $\mb{15.48}$ \\
        %{\it C38} & $15.14$      & $\ul{14.51}$ & $14.93$      & $\mb{14.26}$ \\
        %{\it C41} & $16.18$      & $\ul{15.01}$ & $15.71$      & $\mb{14.95}$ \\
        %{\it C44} & $\ul{12.03}$ & $12.17$      & $12.71$      & $\mb{11.74}$ \\
        \rowcolor[gray]{0.8}%
        Avg. of $42$ & $14.81$ & $\ul{14.04}$ & $14.55$ & $\mb{13.82}$ \\
        \hline
        Avg. Time & $104.28$ & $18.93$ & $\mb{0.90}$ & $\ul{1.08}$ \\
        \hline\hline
        {\it WalkingGirl}   & $\ul{12.92}$ & $13.57$      & $13.65$ & $\mb{12.62}$ \\
        {\it WalkingOnSnow} & $11.06$      & $\ul{10.90}$ & $11.11$ & $\mb{10.60}$ \\
        {\it ICICS}         & $17.60$      & $\ul{17.56}$ & $17.59$ & $\mb{16.40}$ \\
        {\it UBC}           & $\ul{12.54}$ & $13.19$      & $13.29$ & $\mb{12.33}$ \\
        \rowcolor[gray]{0.8}%
        Avg. of $4$ & $\ul{13.53}$ & $13.81$ & $13.91$ & $\mb{12.99}$ \\
        \hline
        Avg. Time & $99.49$ & $8.55$ & $\mb{0.91}$ & $\ul{1.10}$ \\
        \thline
        \end{tabular}
        \label{Tab_BitrateTime}
    \end{table}
    
    \begin{figure}[t]
        \centering
        \includegraphics[scale=0.375,keepaspectratio=true]{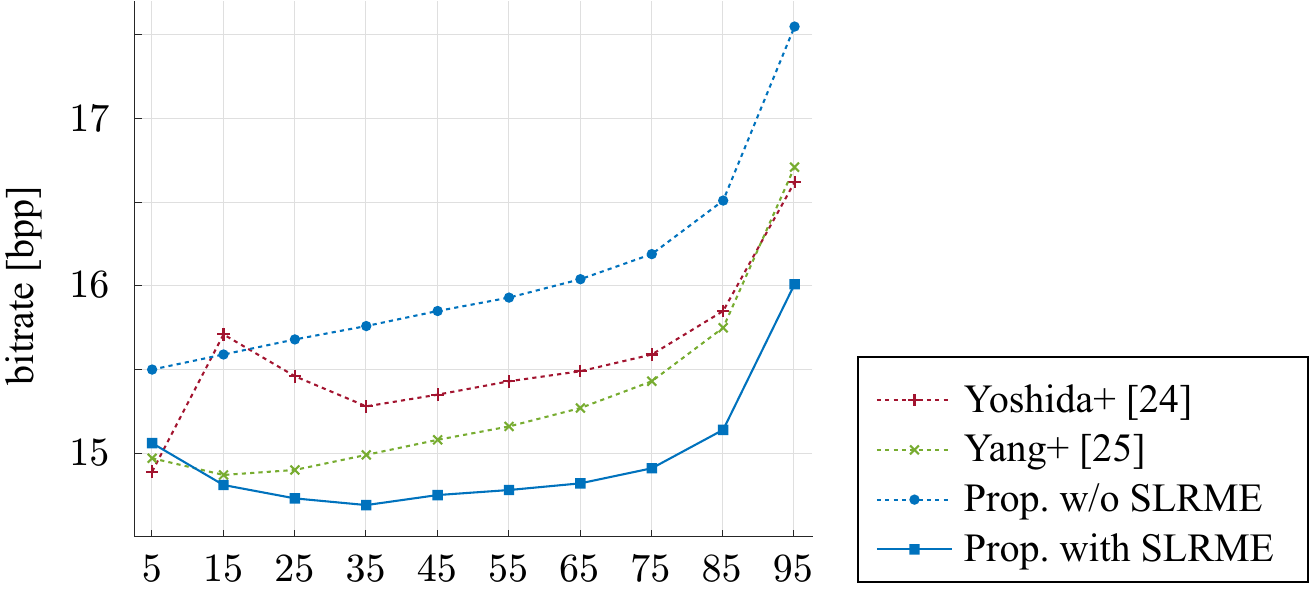}\\
        ~\\
        \includegraphics[scale=0.375,keepaspectratio=true]{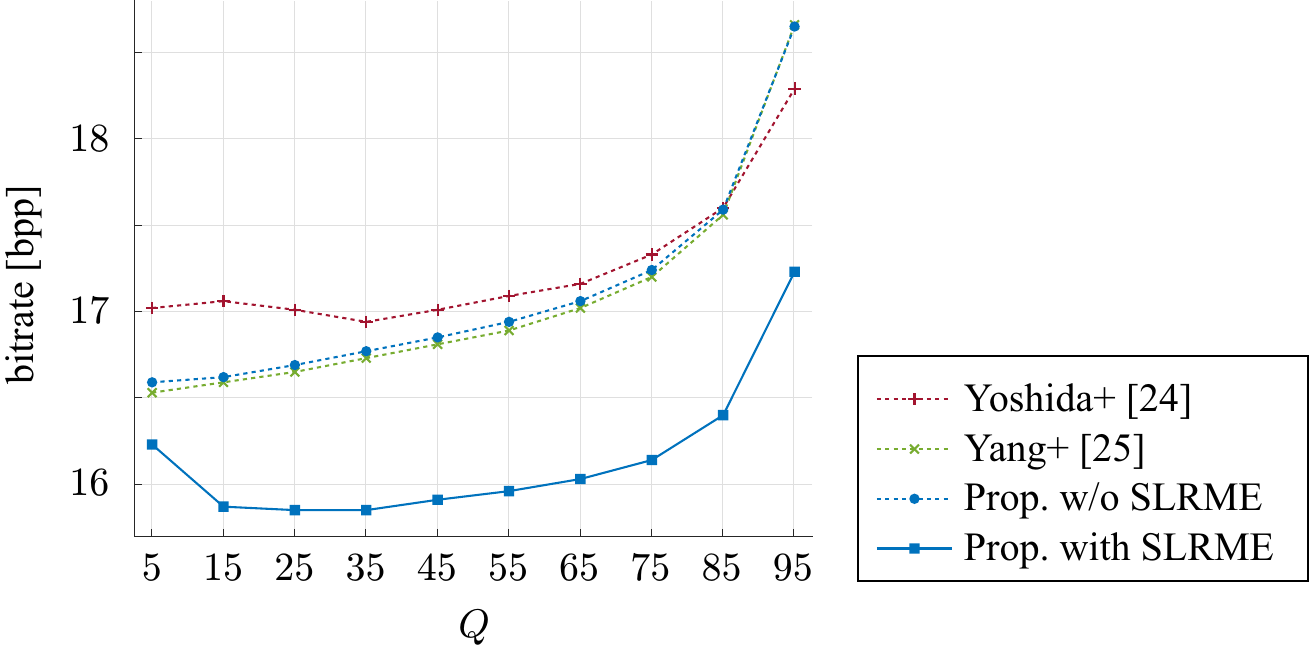}
        \caption{Relationships between $Q$ and bitrate: (top) {\it C23} and (bottom) {\it ICICS}.}
        \label{Fig_QvsBPP}
    \end{figure}

In the experiment, we compared our coding (with/without the SLRME) with two existing dual-layer lossless codings for HDRIs: Yoshida et al.'s method~\cite{Yoshida2018IEICE} and Yang et al.'s method~\cite{Yang2020APSIPA}.
Since we were interested in lossless coding, we excluded from consideration methods that are irreversible.
We used $1920\times 1080$ full-color Radiance HDRIs from the HDR-Eye dataset~\cite{Nemoto2015VPQM} and DML-HDR dataset~\cite{Dehkordi2014QSHINE} and measured the performance on MATLAB (2021b) running on an Intel Core i9-11900K CPU.
We used Reinhard et al.'s method~\cite{Reinhard2002ToG} for the TMO, Huo et al.'s method~\cite{Huo2014VC} for the ITMO,\footnote{Only Yang et al.'s method used the ITMO.} JPEG for the base layer coder (Encoder/Decoder 1 in Fig.~\ref{Fig_Flow}), and JPEG 2000 lossless mode for the enhancement layer coder (Encoder/Decoder 2 and 3 in Fig.~\ref{Fig_Flow}).\footnote{JPEG and JPEG 2000 were implemented by `imwrite.m' in MATLAB.}
While the selection of methods somewhat affects coding performance, the above methods are representative.
We adjusted the residuals to be positive values, as JPEG 2000 does not allow negative input signals, and transmitted the adjusted parameter as extra information.
Finally, all parameters that were transmitted as extra information were considered not to affect the bitrate calculation as the amount of extra information was negligible.

Table~\ref{Tab_BitrateTime} shows examples of bitrates [bits per pixel (bpp)], average bitrates [bpp], and average encoder implementation times [seconds (sec)] of lossless coding for HDRIs with a JPEG quality factor of $Q=85$.
Figure~\ref{Fig_QvsBPP} shows the relationships between $Q$ and bitrate in {\it C23} in \cite{Nemoto2015VPQM} and {\it ICICS} in \cite{Dehkordi2014QSHINE} as representative examples.
Our coding with the SLRME reduced the average bitrate by approximately $5.02$ \% relative to the coding without the SLRME and did so at only a small increase in computational cost.
Moreover, it reduced the average bitrate by approximately $1.57$--$6.68$ \%, while significantly reducing the average encoder implementation time by approximately $87.13$--$98.96$ \% compared with the existing methods.
These results show that SLRME can achieved an excellent level of performance.

%%%%%%%%%%%%%%%%%%%%%%%%%%%%%%%%%%%%%%%%%%%%%%%%%%
\section{Conclusions}
%%%%%%%%%%%%%%%%%%%%%%%%%%%%%%%%%%%%%%%%%%%%%%%%%%
We proposed a lightning-fast dual-layer lossless coding for Radiance HDRIs.
To efficiently suppress the dynamic range of residuals between the HDRI and SDRI in the enhancement layer, our coding directly uses the mantissa and exponent information from the format.
To further reduce the residual energy at a low computational cost, we devised the SLRME for the encoded-decoded SDRI.
Experimental results demonstrated that our coding outperformed existing methods in terms of bitrate, while significantly reducing the encoder implementation time.

\pagebreak
\bibliography{main} %hoge.bibから拡張子を外した名前
\bibliographystyle{IEEEbib} %参考文献出力スタイル

\end{document}